\documentclass[a4paper]{article}
\usepackage[T1]{fontenc}
\pdfoutput=1 

\usepackage{PRIMEarxiv}

\newgeometry{
    textheight=8in,
    top=2cm,
    bottom=2cm,
  }
  
\usepackage{amsmath,amssymb,amsfonts}
\usepackage[hidelinks]{hyperref}

\usepackage{tikz}

\usepackage[justification=centering]{caption}
\usepackage{subcaption}

\newcommand{\sharedir}{share}

\def\E#1{\mathbb{E}\left[#1\right]}

\def\thetab{{\boldsymbol{\theta}}} 
\def\ub{\mathbf{u}} 
\def\Ub{\mathbf{U}} 
\def\yb{\mathbf{y}} 

\def\Rb{\mathbf{R}} 
\def\Kb{\mathbf{K}} 
\def\ub{\mathbf{u}}

\makeatletter
\let\oldtheequation\theequation
\renewcommand\tagform@[1]{\maketag@@@{\ignorespaces#1\unskip\@@italiccorr}}
\renewcommand\theequation{(\oldtheequation)}
\makeatother

\newcommand\anonym[2]{#1}

\begin{document}

\fancyhead{} 


\def\mydoi{<TBD>}
\def\copyrightnotice{\textbf{Note}: Paper accepted at the 5th International Conference on Machine Learning for Networking (MLN'2022) and will be published as a post-proceedings in Springer's LNCS.}

\title{Low Complexity Adaptive Machine Learning Approaches for End-to-End Latency Prediction \thanks{\copyrightnotice}}

\ifx\anonymousPaper\undefined


    \author{
      Pierre Larrenie \\
      Thales SIX \& LIGM \\
      Université Gustave Eiffel, CNRS \\
      Marne-la-Vallée, France\\
      \texttt{pierre.larrenie@esiee.fr} \\
       \And
      Jean-François Bercher \\
      LIGM \\
      Université Gustave Eiffel, CNRS \\
      Marne-la-Vallée, France\\
      \texttt{jean-francois.bercher@esiee.fr} \\
      \AND
      Olivier Venard \\
      ESYCOM \\
      Université Gustave Eiffel, CNRS \\
      Marne-la-Vallée, France\\
      \texttt{olivier.venard@esiee.fr} \\
      \And
       Iyad Lahsen-Cherif \\
      Institut National des Postes et Télécommunications (INPT) \\
      Rabat, Morocco \\
      \texttt{lahsencherif@inpt.ac.ma} \\
    }

\fi

\maketitle              

\begin{abstract}
Software Defined Networks have opened the door to statistical and
AI-based techniques to improve efficiency of networking. Especially to
ensure a certain \emph{Quality of Service} (QoS) for specific
applications by routing packets with awareness on content nature (VoIP,
video, files, etc.) and its needs (latency, bandwidth, etc.) to
use efficiently resources of a network.

Monitoring and predicting various Key Performance Indicators (KPIs) at any level may
handle such problems while preserving network bandwidth.

The question addressed in this work is the design of efficient, 
low-cost adaptive algorithms for KPI estimation, monitoring and prediction. We focus on end-to-end latency prediction, for
which we illustrate our approaches and results on data obtained from a public generator provided after the recent international challenge on GNN \cite{suarez2021graph}. 

In this paper, we improve \anonym{our}{} previously proposed low-cost estimators \cite{larrenie2022icccnt} by adding the adaptive dimension, and show that the performances are minimally modified
while gaining the ability to track varying networks.

\end{abstract}
    \section{Introduction}

Routing while ensuring quality of service (QoS) remains a significant challenge in all networks. Whatever the resources, their use must be optimized to satisfy both throughput and QoS to users.  This is true for static wide area networks, but even more so for mobile networks with dynamic topology. 

The emergence of software-defined networks (SDNs) \cite{singh2017sdnsurvey,amin2018sdnsurvey} has enabled data to be shared more efficiently across communication layers.
Services can provide network requirements to routers; routers acquire data about network performance and allocate resources to meet those requirements as best as possible.
However, acquiring overall network performance can result in high network bandwidth consumption for signaling, degrading the available resources, and is particularly limiting for resource-constrained networks such as mobile networks (MANETs).

We consider a network for which we wish to reduce signaling and perform intelligent routing. In order to limit the amount of signaling, the first axis is to estimate some key performance indicators (KPIs) from other KPIs. A second axis would be to perform this prediction locally, at the node level, rather than a global estimation in the network. Finally, if predictions are to be performed locally, the complexity of the algorithms will need to be low while preserving good prediction quality.  The last point is to be able to detect and track changes in the state of the network, which implies that the predictors will have to use only a small number of the previous states of the network and be able to readapt continuously. 

The question addressed in this work is the design of efficient, low-cost adaptive algorithms for KPI estimation, monitoring and prediction.  
In the present paper, we improve \anonym{our}{} previously proposed low-cost estimators \cite{larrenie2022icccnt} by adding the adaptive dimension and show that the performances are minimally modified 
while gaining the ability to track varying networks.
We focus on end-to-end latency prediction, for which we illustrate our approaches and results on data we generated using a public generator made available after the recent international challenge \cite{suarez2021graph}. 
The best performances of the state-of-the-art are obtained with Graph Neural Networks (GNNs) \cite{rusek2019unveiling,itubnngnn2020,suarez2021graph}. Although this is a global method while we favor local and adaptive methods, we used these performances as a benchmark.

We present related works in \autoref{sec:key_concepts}. Then we present in \autoref{sec:our_ml_approaches} \anonym{our}{the} main results from \cite{larrenie2022icccnt}. Instead of using high performances global and high-costs methods based on Graph Neural Networks (GNNs) \cite{rusek2019unveiling,itubnngnn2020,suarez2021graph}, 
we proposed to use standard machine learning regression methods. We showed that a careful feature engineering and feature selection (based on queue theory and the approach in \cite{parana2021}), as well as the use of a single feature with curve-fitting methods, allows to obtain near state-of-the-art performances with both a very low number of parameters,
significantly lower learning and inference times compared to GNNs, and the with the ability to operate at the link level instead of a whole-graph level.  In \autoref{sec:adaptiveversions}, we show how these block algorithms can be transformed into versions implementable in an iterative way (i.e. by taking into account the data one by one as they become available), with the originality of using a regularization term. Then, time dependent estimations, or the addition of forgetting factor will give them an adaptive character. In \autoref{sec:experimentsAndResults} we describe the validation dataset we built from a public generator and then the results of our experimentation. Finally, we conclude, discuss the overall results and draw some perspectives.

\section{\texorpdfstring{Related work
\label{sec:key_concepts}}{Related work}}


\cite{chua2016stringer} present an heuristic and an Mixed
Integer Programming approach to optimize Service Functions Chain
provisioning when using Network Functions Virtualization for a service
provider. Their approach relies on minimizing a trade-off between the
expected latency and infrastructures resources.

Such optimization routing flow in SDN may need additional information
to be exchanged between the nodes of a network. This results in an
increase of the volume of signalization, by performing some measurements
such as in \cite{pasca2017amps}. This is not a consequent problem in
unconstrained networks, i.e.~static wired networks with near-infinite
bandwidth but may decrease performance of wireless network with poor
capacity. An interesting solution to save bandwidth would be to predict
some of the KPIs from other KPIs and data exchanged globally between
nodes.

In \cite{poularakis2018sdn,poularakis2019tacticalsdn}, authors proposed a MANETs application of SDN in the domain of tactical networks. They proposed a multi-level SDN controllers architecture to build both secure and resilient networking. While orchestrating communication efficiently under military constraints such as: high-level of dynamism, frequent network failures, resources-limited devices.  The proposed architecture is a trade-off between traditional centralized architecture of SDN and a decentralized architecture to meet dynamic in-network constraints. 

\cite{jahromi2018towards} proposed a Quality of Experience (QoE) management strategy in a SDN to optimize the loading time of all the tile of a mapping application. They have shown the impact of several KPIs on their application using a Generalized Linear Model (GLM). This mechanism make the application aware of the current network state.

\begin{figure}[!ht]
    \centering
    \includegraphics[width=2.5in]{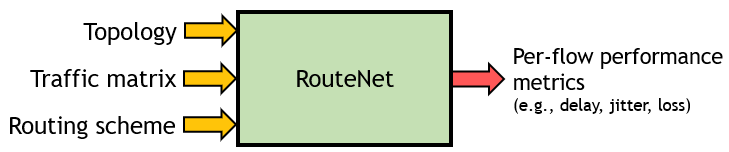}
    \caption{Routenet Architecture \cite{rusek2019unveiling}}
    \label{fig:routenet}
    
\end{figure}

\begin{figure*}\centering
\subfloat[Simple topology]{\begin{tikzpicture}[scale=0.7]
        \begin{scope}[node_topo/.style={circle,fill,draw,text=white,font=\sffamily,minimum
        size=9mm}, edge_topo/.style={thick, stealth-stealth}, flow_topo/.style={darkgray, -stealth, thick, dashed}]
            \node[node_topo] (v1) at (0.5,1) {A};
            \node[node_topo](v4) at (0.5,-2.5) {B};
            \node[node_topo] (v3) at (-1,-1) {C};
            \node[node_topo] (v2) at (2.5,0) {D};
            \node[node_topo] (v5) at (3,-2) {E};
            \draw[thick, stealth-stealth]  (v1) edge (v2);
            \draw[edge_topo]  (v1) edge (v3);
            \draw[edge_topo]  (v4) edge (v3);
            \draw[edge_topo]  (v4) edge (v5);
            \draw[edge_topo]  (v1) edge (v4);
            \draw[edge_topo]  (v2) edge (v5);
            \draw[flow_topo]
                        (-0.8,-1.4) .. controls (0,-3.5)  and (0,-3.5) .. (2.65,-2.2);
            \node[darkgray] (F1) at (0,-3.3) {$F_3$};
            \draw[flow_topo]
                        (0.65,0.60) .. controls (0.65,-2.5)  and (0.355,-2.5)  .. (2.5,-1.9);
            \node[darkgray] (F2) at (1,-1.8) {$F_2$};
            \draw[flow_topo]
                        (1,1) .. controls (3.2,0.6)  and (3.2,0.6)  .. (3.25,-1.5);
            \node[darkgray] (F3) at (3.4,0) {$F_1$};
        \end{scope}
\end{tikzpicture}}\quad
\subfloat[Paths-links Hypergraph of (a)]{\begin{tikzpicture}[scale=0.7]
    \begin{scope}[flow_node/.style={circle,dashed,draw,text=darkgray,font=\sffamily,minimum
            size=10mm},link_node/.style={circle,draw,text=black,font=\sffamily,minimum size=10mm}, link_use/.style={black, -stealth, thick}]
            \node[flow_node] (F1) at (-1,0) {$F_1$};
            \node[flow_node] (F2) at (2,0) {$F_2$};
            \node[flow_node] (F3) at (5,0) {$F_3$};

            \node[link_node] (LAD) at (-1, -2) {$L_{AD}$};
            \node[link_node] (LDE) at (0.5,-2) {$L_{DE}$};
            \node[link_node] (LAB) at (2,-2) {$L_{AB}$};
            \node[link_node] (LBE) at (3.5,-2) {$L_{BE}$};
            \node[link_node] (LCB) at (5,-2) {$L_{CB}$};

            \draw[link_use]  (F1) edge (LAD)
                                edge (LDE);			
            \draw[link_use]  (F2) edge (LAB)
                                edge (LBE);
            \draw[link_use]  (F3) edge (LBE)
                                edge (LCB);
    \end{scope}
\end{tikzpicture}}
\caption{
    Routenet \cite{rusek2019unveiling} paths-links hypergraph transformation applied on a simple topology graph carrying 3 flows. \\
    {\small (a) Black circles represents communication node, double headed arrows between them denotes available symmetric communications links and dotted arrows shows flows path. (b) Circle (resp. dotted) represents links (resp. flows) entities defined in the first graph ($L_{ij}$ is the symmetric link between node $i$ and node $j$.). Unidirectional arrows encode the relation "\textless flow\textgreater~is carried by \textless link\textgreater ".}
}
\label{fig:routenet_hypergraph}

\end{figure*}
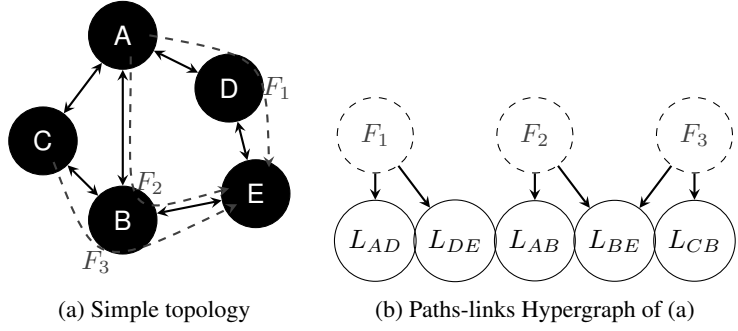

\cite{rusek2019unveiling} used GNNs for predicting
KPIs such as latency, error-rate and jitter. They relied on the
\emph{Routenet} architecture of \autoref{fig:routenet}. The idea is to
model the problem as a bipartite hypergraph mapping flows to links as
depicted on \autoref{fig:routenet_hypergraph}. Aggregating messages in
such graph may result in predicting KPIs of the network in input. The
model needs to know the routing scheme, traffic and links properties.
Their result is very promising and has been the subject of two ITU
Challenge in 2020 and 2021 \cite{itubnngnn2020,suarez2021graph}.
These ITU challenges have very good results since the top-3 teams are around
2\% error in delay prediction in the sense of Mean-Absolute Percentage Error (MAPE).

In \cite{parana2021}, very promising
results were obtained with a a near 1\% GNN model error (in the sense of MAPE)
on the test set.
The model mix analytical \(M/M/1/K\) queueing theory used to create
extra-features to feed GNN model. In order to satisfy the constraint of
scalability proposed by the challenge, the first part of model operates
at the link level.

\section{\label{sec:our_ml_approaches}Simple machine-learning approaches for latency prediction}


Our first problem is to define an estimator $\hat{y}$ of the occupancy $y$ as a function of the different available ``features'' of the system, with a joint objective of low complexity and performance. To do so, we will look for an approximation function $f_\thetab(\ub)$ that allows to estimate $y$ from the features $\ub$ and parameters $\thetab$. 
\begin{equation}
\label{eq:genericmodel}
\hat{y} = f_\thetab(\ub)
\end{equation}
Here $\ub$ and $\thetab$ are vectors that collect the different features or parameters. 
Once an estimate of occupancy is obtained, it is possible to get the latency prediction $\hat{d_n}$ for a specific link $n$ by the simple relation
\begin{equation}
    \hat{d_n} = \hat{y_n}\frac{\mathbb{E}(|P_n|)}{c_n}
\end{equation}
where $\E{|P_n|}$ is the observed average packet size on link $n$ and $c_n$ the capacity of this link.

For analytical simplicity, the parameters $\thetab$ will be sought by minimizing the minimum mean square error
\begin{equation}
\label{eq:genericmmse}
\E{(y - \hat{y})^2} = \E{(y - f_\thetab(\ub))^2} ,
\end{equation}
although the performances are also often evaluated in the MAPE sense
\begin{equation}
    \mathcal{L}\left(\hat{y}, y \right) = \frac{100\%}{N} \sum_{n=1}^N \left|\frac{\hat{y}_n - y_n}{y_n}\right|
\end{equation}
which is preferred to Mean Squared Error (MSE) because of its scale-invariant property. 

We will focus here on two very simple models, although other machine learning models have also been considered in \cite{larrenie2022icccnt}.  Indeed, these two models lend themselves very easily to an adaptive formulation. In this section, we will first describe these two approaches and their performances, before giving the general adaptive formulation, which we will particularize in both cases.

\subsection{\texorpdfstring{Feature Engineering and Linear Regression
\label{sec:feature_engineering}}{Feature Engineering and Machine Learning}}

Based on the assumption that the system may be approximated by a model whose essential features come from \(M/M/1/K\) and \(M/G/1/K\) queue theory, we took essential parameters characterizing queueing systems, such as: $\rho$, $\rho_e$, $\pi_0$, $\pi_K$, etc. and built further features by applying interactions and various non-linearities (powers, log, exponential, square root). Then, we selected features in this set by a forward step-wise selection method; i.e. by adding in turn each feature to potential models and keeping the feature with best performance. Finally, we selected the model with best MAPE error. For a linear regression model, this led us to select and keep a set of 4 simple features, which interestingly enough, have simple interpretations:

\begin{equation}
    \begin{cases}
        \pi_0 = \frac{1-\rho}{1-\rho^{K+1}} \\
        L = \rho + \pi_0\sum_k k\rho^k \\
        \rho_e = \frac{\lambda_e}{\lambda}\rho = \frac{\lambda_e}{\mu} \\
        S_e = \sum_k k\rho_e^k
    \end{cases}\label{eq:mvfeatures}
\end{equation}
where $L$ is the expected number of packets in the queue according to $M/M/1/K$, $\pi_0$ the probability that the queue is empty according to $M/M/1/K$ theory, $\rho_e$ the effective queue utilization, and $S_e$ the unnormalized expected value of the effective number of packet in the queue buffer.
These features can be thought as a kind of data preprocessing, before applying ML algorithms, and this turns out to be a key to achieving good performances. The 4 previous features have been used as input for several machine learning models like Multi-Layer Perceptron model (MLP),
Linear Regression, SVM, Random Forest, Gradient Boosting Regression Tree. We only describe here the case of linear regression, since it is a method for which an adaptive version is readily obtained. In this case, model \eqref{eq:genericmodel} is simply
\begin{equation}
  \label{eq:linearregresionmodel}
    \hat{y} = \theta_0 + \theta_1 \pi_0 + \theta_2 L + \theta_3\rho_e + \theta_4 S_e = \thetab^T \ub
\end{equation}
with $\thetab^T = [\theta_0, \ldots \theta_4]$ and $\ub^T = [1, \pi_0, L, \rho_e, S_e ]$. 
For the linear regression model in (\eqref{eq:linearregresionmodel}, it is well known that the regularized minimum mean squared error 
\begin{equation}
\label{eq:mmselinear}
    J(\thetab) = \E{(y - \thetab^T \ub)^2} + \alpha ||\thetab||^2
\end{equation}
is obtained for 
\begin{equation}
\thetab:  \left(\Rb_{uu} + \alpha \mathbf{1}\right) \thetab= \Rb_{yu} 
\label{eq:normaleq} 
\end{equation}
where we  denoted 
$$
\begin{cases}
\Rb_{uu} = \E{\ub \ub^T}, & \text{the correlation matrix of $\ub$} \\
\Rb_{yu} = \E{y \ub }, & \text{the correlation vector of $y$ and $\ub$} 
\end{cases}
$$
and $\mathbf{1}$ the identity matrix,  $\alpha$ the regularization parameter. 

As far as performance is concerned with this approach, it was evaluated using static data from the GNN ITU Challenge 2021 \cite{suarez2021graph}. Compared to the state-of-the-art, our linear regression with carefully selected features shows a very slight performance degradation: 1.74\% in MAPE while the best state-of-the-art method is at 1.27\%. One strong advantage is in term of training and inference time.  It has a training time of less than a second when GNN requires more than 8 hours. Moreover, the inference time for the complete network is also much lower, by a factor of almost 1000 (0.296s vs 214s).

\subsection{Curve Regression by Bernstein polynomials}
\label{sec:curveregression}

\begin{figure}[!ht]
    \centering
    \includegraphics[width=3in]{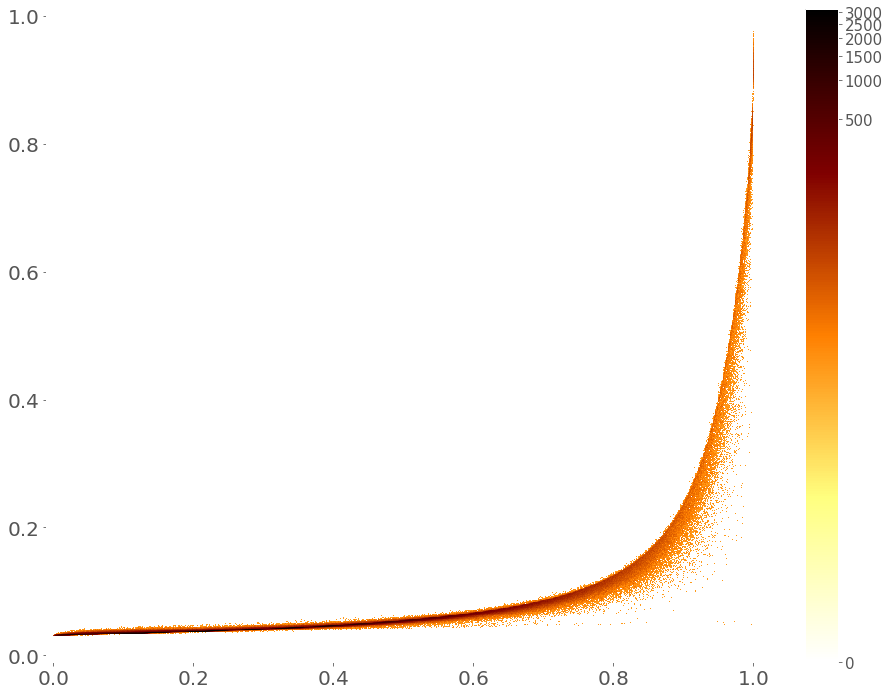}
    \caption{Data of ITU Challenge 2021 \cite{suarez2021graph}, $\rho_e$ vs queue occupancy. Color-scale is an indicator of points cloud density.    }
    \label{fig:datarhoy}
\end{figure}

There is a high interdependence of the features we selected in 
\autoref{eq:mvfeatures}, since all these features can be expressed in term of \(\rho_e\).
Furthermore, it is confirmed by data exploration that $\rho_e$ is the prominent feature for occupancy prediction (and in turn latency prediction), as exemplified in \autoref{fig:datarhoy}. 

It is then tempting to try to further simplify our features space and estimate the occupancy from a non-linear transformation of the single feature $\rho_e$, as:
\begin{equation}
    \hat{y} = g(\rho_e)
\end{equation} 
where \(\hat{y}\) is the estimate of the occupancy \(y\). 
The concerns are of course to define simple and efficient functions $g$, with a low number of parameters, that can model the  kind of growth shown in \autoref{fig:datarhoy}, and of course to check that the performance remains interesting. 


The estimator $g$ is defined as a linear combination of simple functions \(f_n\):
\begin{equation}
\label{eq:polynomialapprox}
    \hat{y}=g(\rho_e)=\sum_n \theta_n \cdot f_n(\rho_e)
\end{equation}
which is also a linear model in terms of function $f_n(\rho_e)$. 

Several solutions were considered in \cite{larrenie2022icccnt} to define or choose the functions \(f_n\). Since we know that the Bernstein polynomials form a basis in the set of polynomial in the interval \([0;1]\); and that the approximation of any continuous function on \([0;1[\) by a Bernstein polynomial converges uniformly, we were led to these polynomials:
\begin{equation}
    f^K_n(x) = \binom{K}{n} x^n(1-x)^{K-n}
\end{equation}
where $K$ is maximum order of polynomials. 

As mentioned, \eqref{eq:polynomialapprox} can be rewritten as the linear model
\begin{equation}
\label{eq:polynomialapproxaslm}
    \hat{y}=g(\rho_e)=\sum_n \theta_n \cdot f_n(\rho_e) = \thetab^T \ub
\end{equation}
with $\thetab^T = [\theta_0, \ldots, \theta_K]$ and $\ub^T = [f^K_0(\rho_e), f^K_1(\rho_e), \ldots, f^K_K(\rho_e) ]$. 
Hence, we have the same form as in \eqref{eq:normaleq} for the solution. 

In term of performances, we also obtained a minor degradation in MAPE (1.68\%) compared to state-of-the-art (1.29\%), while improving by several orders the wall training and inference times (2min/3.14s vs 8hrs/214s); though a bit less than the simple linear regression. 


\section{Adaptive versions}
\label{sec:adaptiveversions}

We place ourselves in the context where we have regular snapshots of the state of the network, which allows us to both monitor the quality of predictions, and to track changes in the network. For the $n$-th series of measurements, let us denote $y(n)$ the measured latency and $\ub(n)$ the features. We can also group several snapshots or several links into a vector of latencies $\yb(n)$ and matrix $\Ub(n)$. In the following we will derive equations for this block case, which includes immediately the scalar case. 

 The minimum mean square error \eqref{eq:mmselinear} which has the explicit solution \eqref{eq:normaleq} can also be solved by a gradient algorithm as
\begin{align}
\thetab_{k+1} & =  \thetab_{k} -\mu \left.\nabla J(\thetab)\right|_{\thetab=\thetab_{k}}, \\
  & =  \thetab_{k} -\mu \left(\left(\Rb_{uu} + \alpha\mathbf{1}\right) \thetab_{k} -  \Rb_{yu}\right).
\label{eq:grad_algo}
\end{align}
%
In \eqref{eq:grad_algo}, we can substitute the true values with estimated ones. 
In order to introduce adaptivity to context changes in the network, these estimates will preserve the temporal dimension. We thus use either a sliding average 
\begin{equation}
\begin{cases}
\hat\Rb_{uu}(n) = \sum_{l=0}^L \Ub(n-l)\Ub(n-l)^T \\
\hat\Rb_{yu}(n) = \sum_{l=0}^L \Ub(n-l)\yb(n-l) 
\end{cases}
\end{equation}
or an exponential mean
\begin{equation}
\begin{cases}
\hat\Rb_{uu}(n) = \sum_{l=0}^n \lambda^{l-n} \Ub(l)\Ub(l)^T = \lambda \hat\Rb_{uu}(n-1) + \Ub(n)\Ub(n)^T\\
\hat\Rb_{yu}(n) = \lambda \hat\Rb_{yu}(n-1) + \Ub(n)\yb(n).\\
\end{cases}
\label{eq:exponentialmean}
\end{equation}
where $\lambda \le 1$ is the forgetting factor. 

In the limit case where we take either $L=0$ or $\lambda=0$ in the previous formulas, we get the `instantaneous estimates`  
\begin{equation}
\begin{cases}
\hat\Rb_{uu}(n) =  \Ub(n)\Ub(n)^T\\
\hat\Rb_{du}(n) =  \Ub(n)\yb(n).\\
\end{cases}
\end{equation}
we obtain 
\begin{equation}
\thetab(n+1) = (1-\mu\alpha)\thetab(n) - \mu \Ub(n) \left(\Ub(n)^T \thetab(n) -  \yb(n) \right) 
\label{eq:lms}
\end{equation}
which reduces to the well known LMS algorithm \cite{widrow1985adaptive} in the scalar case and no regularization, $\alpha=0$. 

Alternatively, one can try to solve the normal equation \eqref{eq:normaleq},  using the time dependent estimates as the exponential mean \eqref{eq:exponentialmean}. The difficulty with the solution
\begin{equation}
\hat\thetab(n+1) =  \left[\hat\Rb_{uu}(n+1) + \alpha\mathbf{1}\right]^{-1} \hat\Rb_{yu}(n+1)
\end{equation}
is the inversion, for each $n$, of the correlation matrix. Let us denote 
\begin{equation}
\Kb({n+1}) = \left[\hat\Rb_{uu}(n+1) + \alpha\mathbf{1}\right]^{-1}.
\end{equation}
Using \eqref{eq:exponentialmean}, we have 
\begin{align}
    \Kb({n+1})^{-1} &= \lambda \hat\Rb_{uu}(n) + \Ub(n+1)\Ub(n+1)^T +  \alpha\mathbf{1} \\
    & = \lambda\left(\hat\Rb_{uu}(n) + \alpha\mathbf{1} \right) + \Ub(n+1)\Ub(n+1)^T + \alpha(1-\lambda)\mathbf{1} \\
    & = \lambda\Kb({n})^{-1}  + \Ub(n+1)\Ub(n+1)^T + \alpha(1-\lambda)\mathbf{1}
\end{align}
and 
\begin{align}
    \Kb({n+1}) &= \left[\left(\lambda\Kb({n})^{-1}  + \Ub(n+1)\Ub(n+1)^T\right) + \alpha(1-\lambda)\mathbf{1}\right]^{-1} \\
    & = \left[\mathbf{Q}(n+1) + \delta \mathbf{1}\right]^{-1}
\end{align}
with 
\begin{equation}
    \mathbf{Q}(n+1) = \left(\lambda\Kb({n})^{-1}  + \Ub(n+1)\Ub(n+1)^T\right)
\end{equation}
and $\delta = \alpha(1-\lambda)$
The matrix inversion lemma enables to reduce the inversion of $\mathbf{Q}(n+1)$ to
 \begin{equation}
 \label{eq:invwood}
    \begin{split}
  \mathbf{Q}(n+1)^{-1} = & 
    \frac{1}{\lambda}\Kb(n)  -
   \frac{1}{\lambda^2} \Kb(n)\Ub(n+1) \times \\   & \left(\mathbf{1}+\frac{1}{\lambda}\Ub(n+1)^T\Kb(n)\Ub(n+1) \right)^{-1} \Ub(n+1)^T \Kb(n),
\end{split}
\end{equation}
which simplifies to
\begin{equation}
\mathbf{Q}(n+1)^{-1} = \frac{1}{\lambda}\Kb(n) - 
\frac{1}{\lambda^2} \frac{\Kb(n)\ub(n+1)\ub(n+1)^T \Kb(n)}{1+\frac{1}{\lambda}\ub(k+1)^T\Kb(n)\ub(k+1)},
\end{equation}
for scalar observations. 
Now, we can use the Taylor expansion to get
\begin{equation}
\label{eq:matrixtaylor}
    \Kb({n+1}) = \left[\mathbf{Q}(n+1) + \delta\mathbf{1} \right]^{-1} = \mathbf{Q}(n+1)^{-1} - \delta \mathbf{Q}(n+1)^{-2} + \delta^2\mathbf{Q}(n+1)^{-3} + \ldots
\end{equation} 
This gives us a way  to compute recursively the inverse of the regularized estimate of the correlation matrix by combining \eqref{eq:invwood} and \eqref{eq:matrixtaylor} into 
\begin{equation}
  \Kb(n+1) \approx \mathbf{Q}(n+1)^{-1} - \delta \mathbf{Q}(n+1)^{-2}
\end{equation}
which, by \eqref{eq:invwood}, does not require the inversion of $K(n)$. 

In both cases, we have the updating formula
\begin{equation}
\thetab(n+1) = \thetab(n) + \Kb(n+1)\Ub(n+1)[\yb(n+1)-\thetab(n)^T\Ub(n+1)].
\end{equation}

\section{Experiments and results}
\label{sec:experimentsAndResults}
\subsection{Dataset}

We generate a dataset thanks to a public challenge data generator \cite{suarez2021graph}. This data generator is based on the well-known OMNET++ discrete event network simulator\cite{omnetpp}. The published simulator is available as a docker image. However, due to the rules of the 2022 edition of the challenge, it is not possible to generate large topologies, i.e. no more than 10 nodes. Since our models are link-based, the use of small topologies does not seem to be a problem. The simulator is parameterized by a traffic matrix and a topological graph that are easy to generate thanks to the provided API.

\begin{figure}[!htbp]
    \centering
    \subfloat[Capacity alteration to model jamming with a decrease of the capacity up to a factor of 5.\label{fig:link_sigmoid}]{%
        \includegraphics[width=\textwidth]{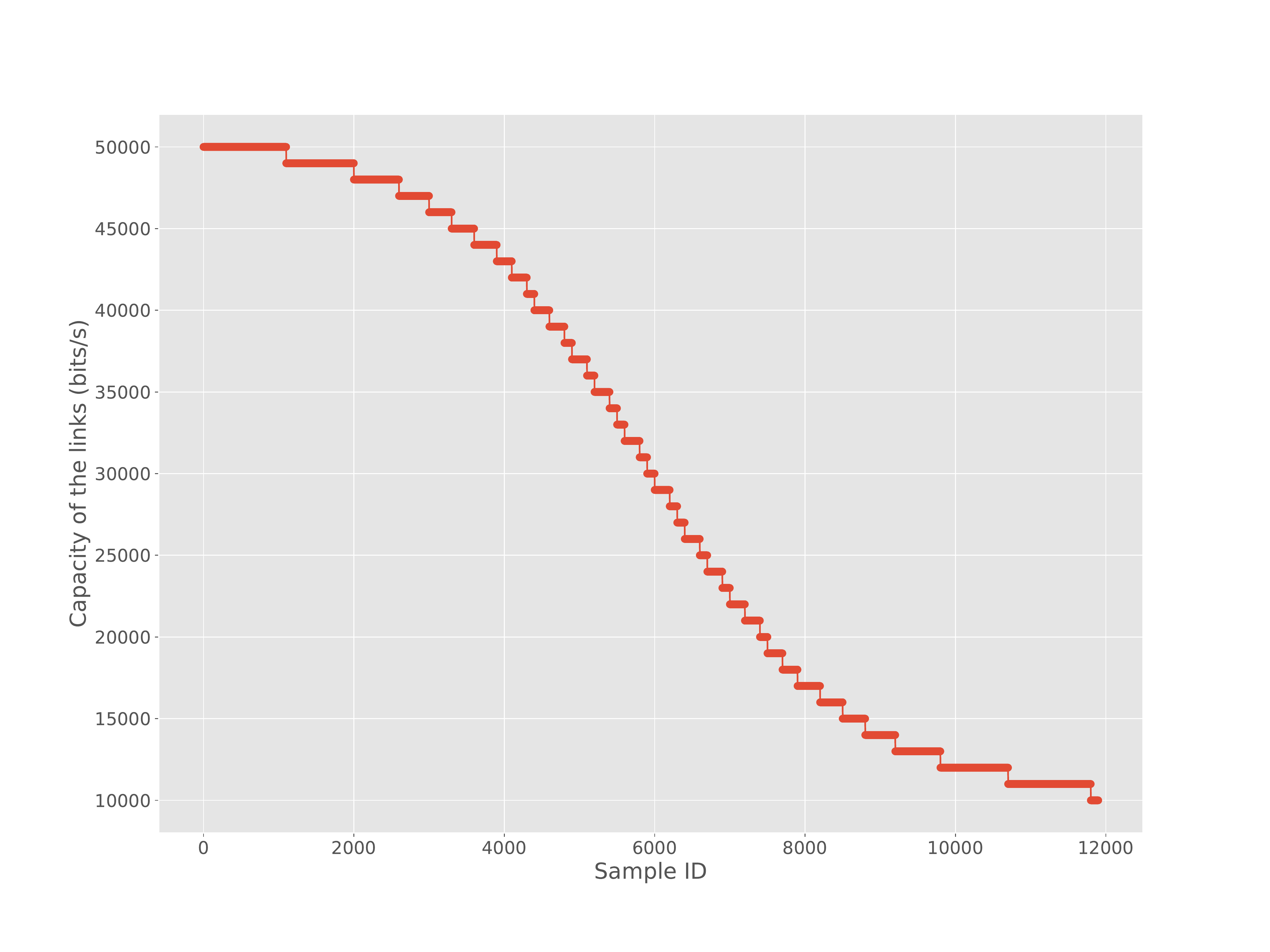}%
    }\hfill
    \subfloat[Link capacity distribution of the generated dataset.\label{fig:data_link_distribution}]{%
        \includegraphics[width=\textwidth]{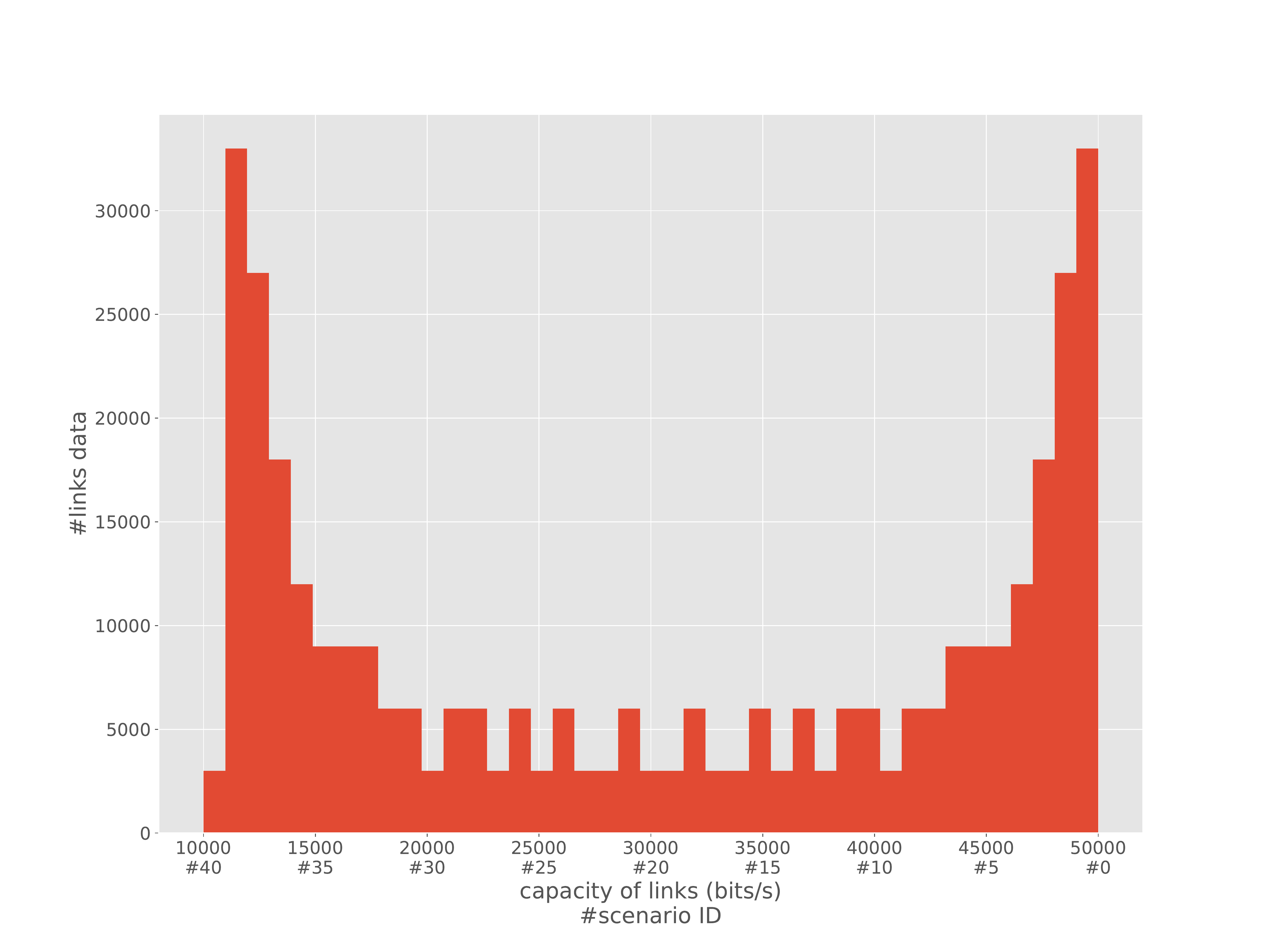}%
    }
    \caption{Overview of the generated dataset}
    \label{fig:overview_generated_data}
\end{figure}

Our generated dataset, used is this paper, is the result of 11900 simulations of the same topology graph of 10 nodes and 30 links, subject to 100 different traffic matrices. In order to get complex results of simulations but at low cost, we made the choice to model a network with small queue buffers (8000 bits) and possibly subject of high traffic intensities (maximum traffic rate set to 4000 bits/s for each flow). Then for each traffic matrices, we alter the capacity of the network according to a sigmoid, in order to model a network subject to jamming, with 2 stationary states. The proposed jamming may cause a decrease in the capacity of the network links by up to a factor of 5, as depicted on \autoref{fig:link_sigmoid}. For simplification purposes, we have considered that each link of the network has the same capacity. This result in a U-shaped distribution of our link data samples according to the link capacity as shown in \autoref{fig:data_link_distribution}. 

\subsection{Results}

From the generated data, we validate our approach along several axes. 
\subsubsection{Global performances}
First, we establish the benchmark performances based on the global methods presented in \anonym{the paper}{} \cite{larrenie2022icccnt} and Sections \ref{sec:feature_engineering} and \ref{sec:curveregression}.

For the linear regression method described in Section \ref{sec:feature_engineering}, we obtain an MSE of 5.86e-4 and a MAPE of 9.58\% for the queue occupancy prediction and an MSE of 1.10e-3 and a MAPE of 10.26\%. for the end-to-end latency prediction.

Concerning the curve regression using Bernstein polynomials (of degree 8) described in in Section \ref{sec:curveregression}, we obtain an MSE of 4.52e-4 and a MAPE of 8.72\% for the queue occupancy prediction and an MSE of 9.35e-4 and a MAPE of 9.95\% for the end-to-end latency prediction.

Note that these benchmark performances are below the performances obtained in \cite{larrenie2022icccnt}, but the dataset we use here is probably more severe since using ground truth value for occupancy results in an MSE 6.03e-4 of a MAPE 9.34\% for the flow delay prediction. That is indeed very close of the obtained results.

\subsubsection{Behavior of iterative algorithms}
\begin{figure}[!htbp]
\centering
\subfloat[Iterative curve-fitting based on Bernstein polynomials of degree 8.\label{fig:global_b8_na}]{%
    \includegraphics[width=\textwidth]{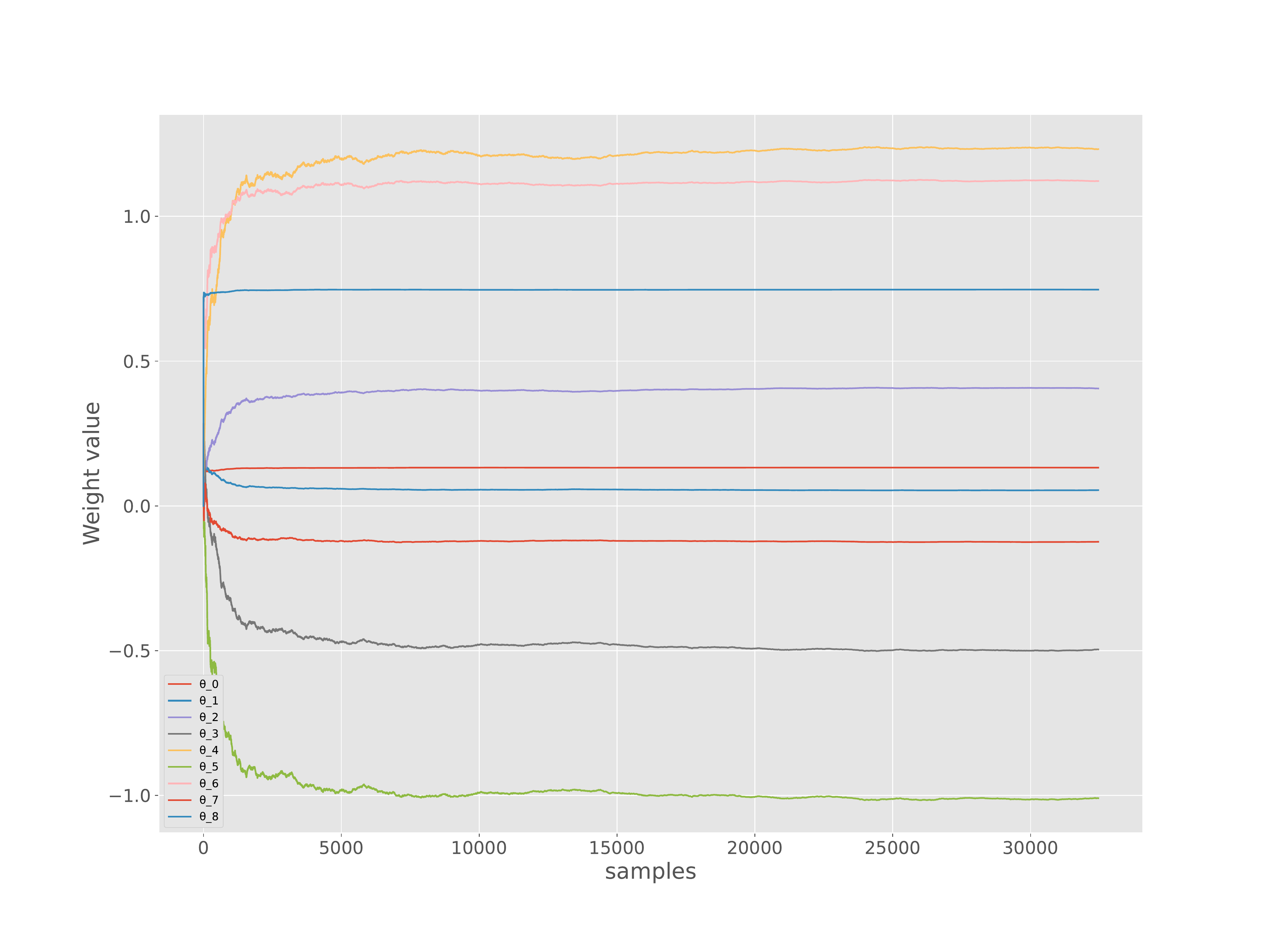}%

}\hfill
\subfloat[Iterative Linear Regression\label{fig:global_lr_na}]{%
    \includegraphics[width=\textwidth]{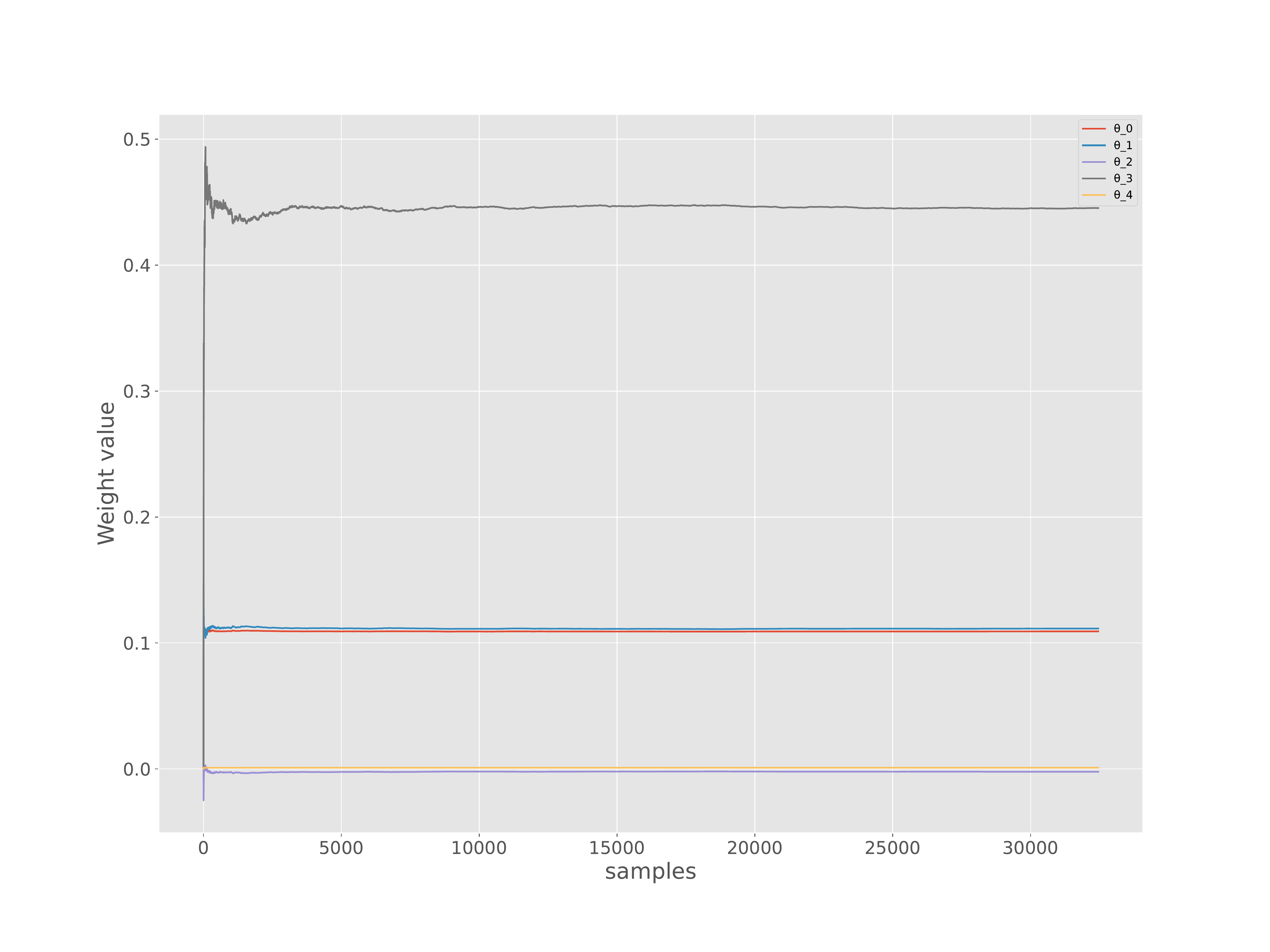}%

}

\caption{Evolution of weights for our iterative methods without forgetting (non-adaptive) while fitting the whole dataset.}
\label{fig:global_iterative_method}

\end{figure}

In a second step, we verify that the algorithms presented in section \ref{sec:adaptiveversions} converge and allow us to recover these performances. With a forgetting factor of 1 (use of all data with the same weight) and a block size of 10, we observe, for example in \autoref{fig:global_iterative_method}, that the model coefficients converge towards a stable value, and that MAPE recovers the value obtained with the global method using all data. The convergence is obtained in less than 10,000 operations. It is thus possible to replace the global method, which is already low-cost, with an approach where the calculations are carried out recursively.  

\subsubsection{Adaptivity}
\begin{figure}[!htbp]
\centering
\subfloat[Evolution of weights along the scenario.\label{fig:ada_bernstein_w}]{%
    \includegraphics[width=\textwidth]{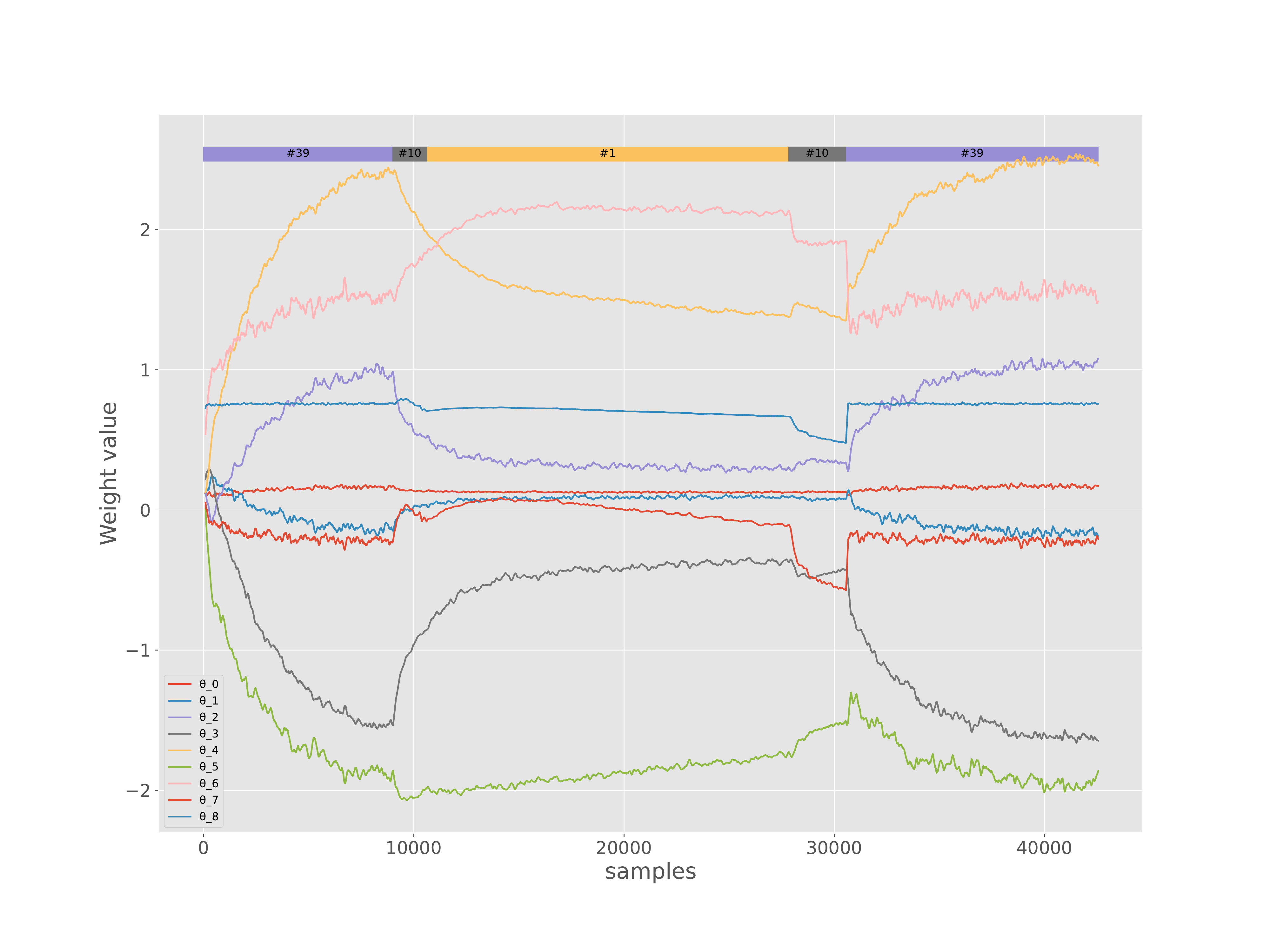}%
}\hfill
\subfloat[Evolution of the RMSE along the scenario.\label{fig:ada_bernstein_err}]{%
    \includegraphics[width=\textwidth]{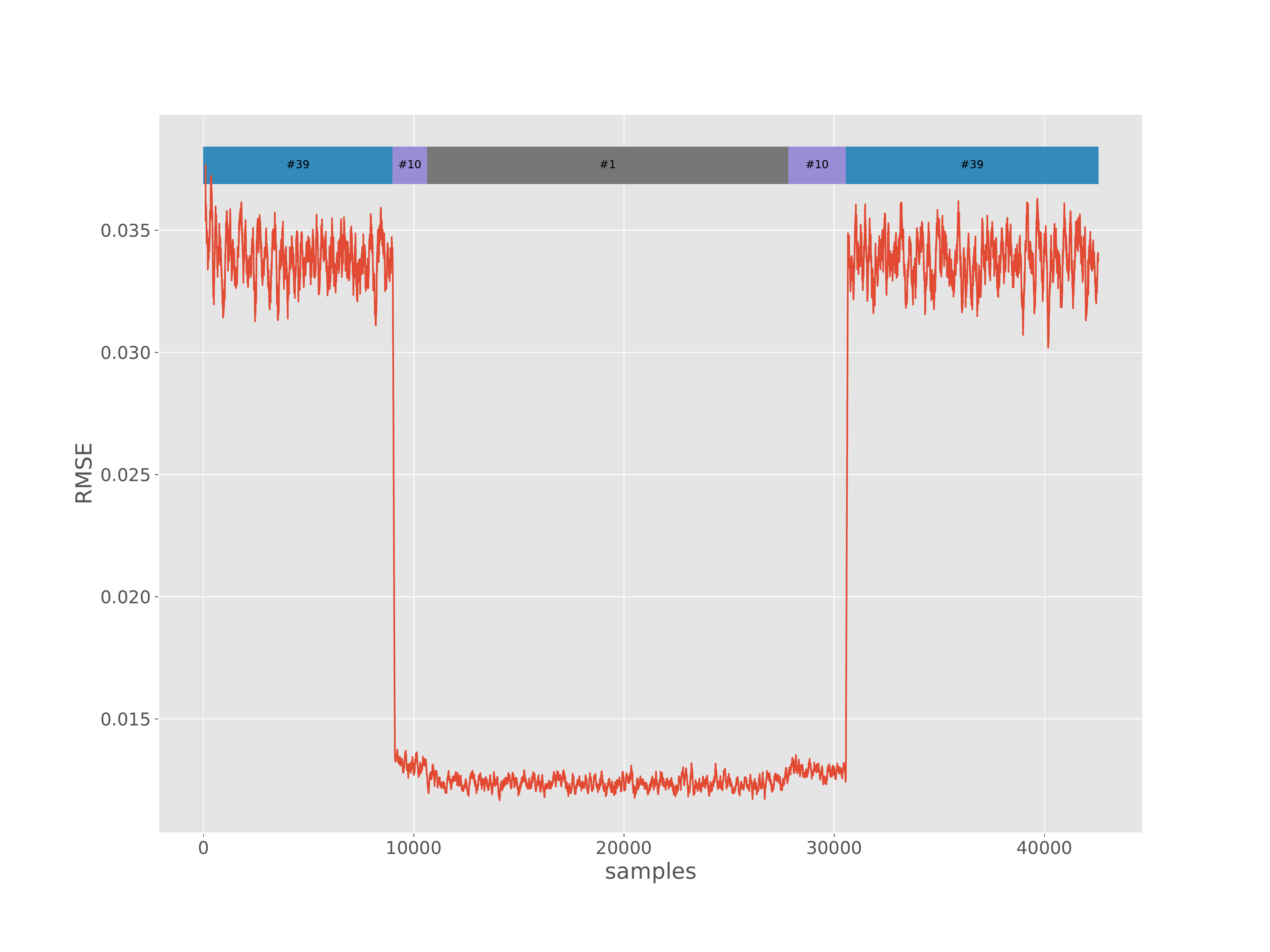}%
}
\caption{Evolution of weights and $\sqrt{\text{MSE}}$ (RMSE) for our adaptive approach of Bernstein polynomial curve regression of degree 8,$\lambda=0.9$, $\alpha=0.08$. Scenario describes a nominal period between 2 periods of jamming. \#39 corresponds to a link capacity of 11 Kbits/s, \#10 40 Kbits/s and \#1 49 Kbts/s. Figure is smoothed over 100 points.}
\label{fig:ada_bernstein}
\end{figure}

\begin{figure}[!htbp]
\centering
\subfloat[Evolution of weights along the scenario.\label{fig:ada_lr_w}]{%
    \includegraphics[width=\textwidth]{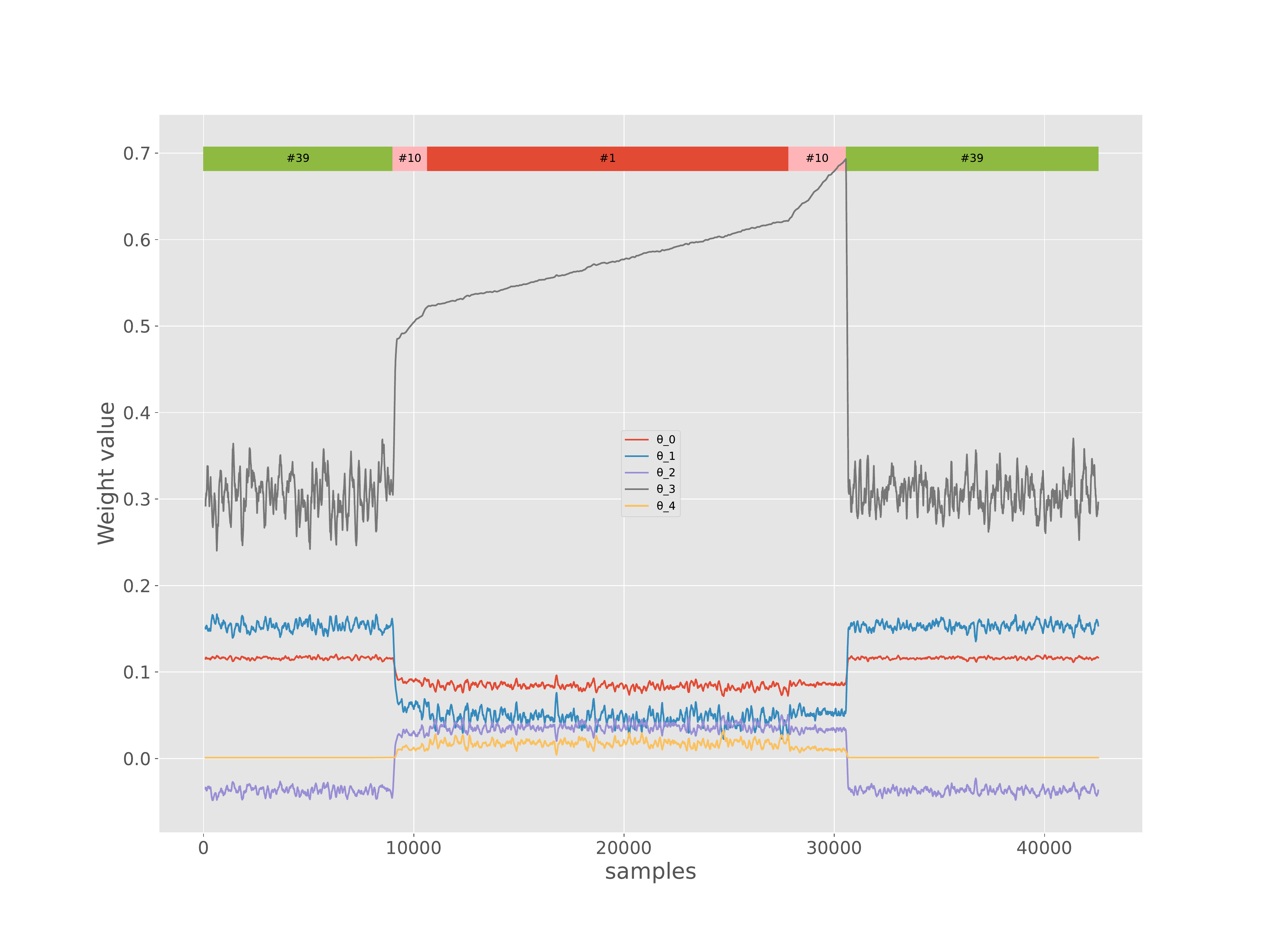}%
}\hfill
\subfloat[Evolution of the RMSE along the scenario.\label{fig:ada_lr_err}]{%
    \includegraphics[width=\textwidth]{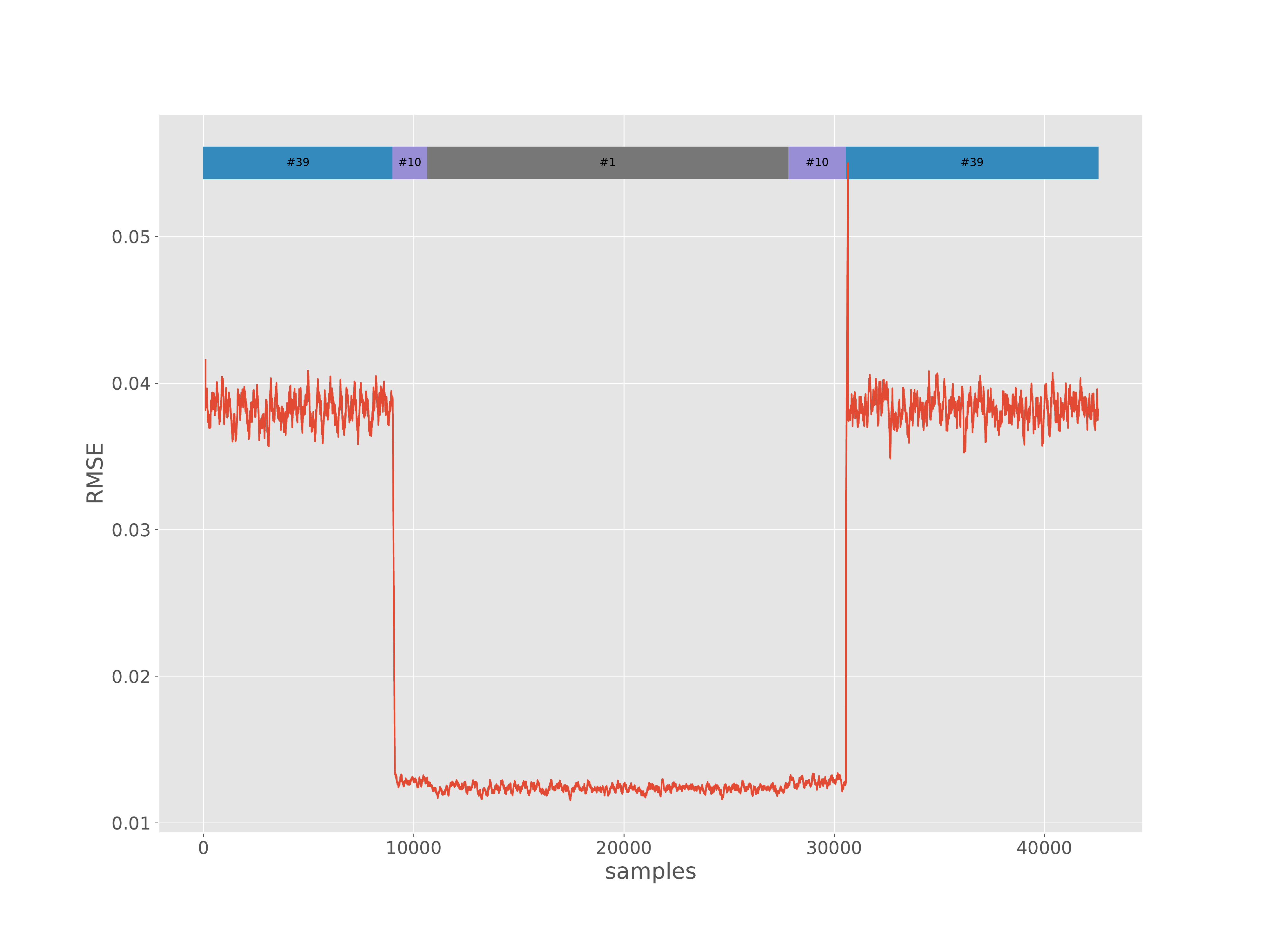}%

}
\caption{Evolution of weights and $\sqrt{\text{MSE}}$ (RMSE) for our adaptive approach of Linear Regression,$\lambda=0.9$, $\alpha=0.08$. Scenario describes a nominal period between 2 periods of jamming. \#39 corresponds to a link capacity of 11 Kbits/s, \#10 40 Kbits/s and \#1 49 Kbts/s. Figure is smoothed over 100 points.}
\label{fig:ada_lr}
\end{figure}

In a third step, we compare the algorithms to the case of network modifications We consider an abrupt change in the network capacity, which could correspond to a jamming scenario. We then examine how the two adaptive algorithms presented (linear regression with judiciously chosen features; and Bernstein polynomial model) can detect and adapt to these modifications. In this context, we examine the role of the forgetting factor and the regularization parameter. \autoref{fig:ada_bernstein} and \autoref{fig:ada_lr} present the results for the case of a capacity change. We observe that (i) the square of the residual error, smoothed over 100 points, is a remarkable indicator of a change in the network; and (ii) that the model coefficients readjust over the iterations after this change. 

\subsubsection{Discussion}

These experiments show the effectiveness and relevance of our iterative and adaptive versions of end-to-end latency estimation procedures. The iterative versions have the same performance as their global counterparts; an even lower cost since they can be implemented iteratively as the data is received or made available. The convergence time for the model coefficients is a few thousand samples while the global model used around 350,000 samples for training, while the GNN models require several million samples. Moreover, we observe that the residual error converges very quickly, in some tens of samples, which means that although the convergence of the models' coefficients does not seem to be complete, they are equivalent from the point of view of performance for occupancy prediction. From an operational point of view, the model can be refreshed regularly, and the predicted KPIs between these updates can be used for intelligent routing. As we have observed, residual error monitoring is an excellent indicator of changes in the network state.

\section{Conclusion}

In this paper, we considered the problem of designing efficient and low-cost algorithms for KPI prediction that are locally implementable and adaptive to network changes. 
Based on a previous work, we have argued and developed adaptive solutions, introducing in addition a regularization term in order to stabilize the results. We used a public domain simulator to simulate networks and generate relevant data. The experiments demonstrate the effectiveness and relevance of these algorithms. 
Thus, we now have low-complexity models that can be implemented iteratively at the level of local links. We have the possibility to predict the occupancy of the different links, and the end-to-end latencies (the models predict the occupancy of the queues, then compute analytically the delay for each link and finally aggregate along the path). Moreover, the adaptability of the solution allows to follow changes in the network state, always at a minimal cost, by re-adapting from the current solution and new data. 
The continuation of the work will focus on the choice criteria of the forgetting factor, on the impact of the regularization factor, in order to find automatic selection methods. Of course, the approaches considered here will have to be considered and adapted for other types of KPI, such as error rate or jitter.

    \newpage
    \bibliographystyle{splncs04}
    \bibliography{bibliography} 

\begin{thebibliography}{10}
\providecommand{\url}[1]{\texttt{#1}}
\providecommand{\urlprefix}{URL }
\providecommand{\doi}[1]{https://doi.org/#1}

\bibitem{amin2018sdnsurvey}
Amin, R., Reisslein, M., Shah, N.: Hybrid {SDN} networks: A survey of existing
  approaches. IEEE Communications Surveys \& Tutorials  \textbf{20}(4),
  3259--3306 (2018)

\bibitem{parana2021}
de~Aquino~Afonso, B.K.: {GNNet} challenge 2021 report (1st place).
  \url{https://github.com/ITU-AI-ML-in-5G-Challenge/ITU-ML5G-PS-001-PARANA}
  (2021)

\bibitem{itubnngnn2020}
{Barcelona Neural Networking Center}: The graph neural networking challenge
  2020. \url{https://bnn.upc.edu/challenge/gnnet2020}

\bibitem{chua2016stringer}
Chua, F.C., Ward, J., Zhang, Y., Sharma, P., Huberman, B.A.: Stringer:
  Balancing latency and resource usage in service function chain provisioning.
  IEEE Internet Computing  \textbf{20}(6),  22--31 (2016)

\bibitem{jahromi2018towards}
Jahromi, H.Z., Hines, A., Delanev, D.T.: Towards application-aware networking:
  Ml-based end-to-end application {KPI/QoE} metrics characterization in {SDN}.
  In: 2018 Tenth International Conference on Ubiquitous and Future Networks
  (ICUFN). pp. 126--131. IEEE (2018)

\bibitem{larrenie2022icccnt}
Larrenie, P., Bercher, J.F., Lahsen-Cherif, I., Venard, O.: Low complexity
  approaches for end-to-end latency prediction. In: Proceedings of the 13th
  IEEE International Conference On Computing, Communication and Networking
  Technologies. IEEE (2022)

\bibitem{pasca2017amps}
Pasca, S.T.V., Kodali, S.S.P., Kataoka, K.: {AMPS}: Application aware multipath
  flow routing using machine learning in {SDN}. In: 2017 Twenty-third National
  Conference on Communications (NCC). pp.~1--6. IEEE (2017)

\bibitem{poularakis2018sdn}
Poularakis, K., Iosifidis, G., Tassiulas, L.: {SDN}-enabled tactical ad hoc
  networks: Extending programmable control to the edge. IEEE Communications
  Magazine  \textbf{56}(7),  132--138 (2018)

\bibitem{poularakis2019tacticalsdn}
Poularakis, K., Qin, Q., Nahum, E.M., Rio, M., Tassiulas, L.: Flexible {SDN}
  control in tactical ad hoc networks. Ad Hoc Networks  \textbf{85},  71--80
  (2019)

\bibitem{rusek2019unveiling}
Rusek, K., Su{\'a}rez-Varela, J., Mestres, A., Barlet-Ros, P.,
  Cabellos-Aparicio, A.: Unveiling the potential of graph neural networks for
  network modeling and optimization in {SDN}. In: Proceedings of the 2019 ACM
  Symposium on SDN Research. pp. 140--151 (2019)

\bibitem{singh2017sdnsurvey}
Singh, S., Jha, R.K.: A survey on {Software Defined Networking}: Architecture
  for next generation network. Journal of Network and Systems Management
  \textbf{25}(2),  321--374 (2017)

\bibitem{suarez2021graph}
Su{\'a}rez-Varela, J., et~al.: The graph neural networking challenge: a
  worldwide competition for education in {AI/ML} for networks. ACM SIGCOMM
  Computer Communication Review  \textbf{51}(3),  9--16 (2021)

\bibitem{omnetpp}
Varga, A., Hornig, R.: An overview of the omnet++ simulation environment. In:
  1st International ICST Conference on Simulation Tools and Techniques for
  Communications, Networks and Systems (2010)

\bibitem{widrow1985adaptive}
Widrow, B., Stearns, S.: Adaptive Signal Processing. Edited by Alan V.
  Oppenheim, Prentice-Hall (1985)

\end{thebibliography}

\end{document}